\documentstyle[preprint,aps,epsfig]{revtex}
\begin{document}

%
\author{
B.~Abbott,$^{48}$                                                             
M.~Abolins,$^{45}$                                                            
V.~Abramov,$^{21}$                                                            
B.S.~Acharya,$^{15}$                                                          
D.L.~Adams,$^{55}$                                                            
M.~Adams,$^{32}$                                                              
V.~Akimov,$^{19}$                                                             
G.A.~Alves,$^{2}$                                                             
N.~Amos,$^{44}$                                                               
E.W.~Anderson,$^{37}$
% ...
\\                                                                            
\vskip 0.30cm                                                                 
\centerline{(D\O\ Collaboration)}                                             
\vskip 0.30cm                                                                 
}                                                                             
\address{                                                                     
\centerline{$^{1}$Universidad de Buenos Aires, Buenos Aires, Argentina}       
\centerline{$^{2}$LAFEX, Centro Brasileiro de Pesquisas F{\'\i}sicas,         
                  Rio de Janeiro, Brazil}                                     
\centerline{$^{3}$Universidade do Estado do Rio de Janeiro,                   
                  Rio de Janeiro, Brazil}                                     
\centerline{$^{4}$Institute of High Energy Physics, Beijing,                  
                  People's Republic of China}                                 
\centerline{$^{5}$Universidad de los Andes, Bogot\'{a}, Colombia}             
\centerline{$^{6}$Charles University, Prague, Czech Republic}                 
\centerline{$^{7}$Institute of Physics, Academy of Sciences, Prague,          
                  Czech Republic}                                             
\centerline{$^{8}$Universidad San Francisco de Quito, Quito, Ecuador}         
\centerline{$^{9}$Institut des Sciences Nucl\'eaires, IN2P3-CNRS,             
                  Universite de Grenoble 1, Grenoble, France}                 
\centerline{$^{10}$CPPM, IN2P3-CNRS, Universit\'e de la M\'editerran\'ee,     
                  Marseille, France}                                          
\centerline{$^{11}$LPNHE, Universit\'es Paris VI and VII, IN2P3-CNRS,         
                  Paris, France}                                              
\centerline{$^{12}$DAPNIA/Service de Physique des Particules, CEA, Saclay,    
                  France}                                                     
\centerline{$^{13}$Panjab University, Chandigarh, India}                      
\centerline{$^{14}$Delhi University, Delhi, India}                            
\centerline{$^{15}$Tata Institute of Fundamental Research, Mumbai, India}     
\centerline{$^{16}$Seoul National University, Seoul, Korea}                   
\centerline{$^{17}$CINVESTAV, Mexico City, Mexico}                            
\centerline{$^{18}$Institute of Nuclear Physics, Krak\'ow, Poland}            
\centerline{$^{19}$Institute for Theoretical and Experimental Physics,        
                   Moscow, Russia}                                            
\centerline{$^{20}$Moscow State University, Moscow, Russia}                   
\centerline{$^{21}$Institute for High Energy Physics, Protvino, Russia}       
\centerline{$^{22}$Lancaster University, Lancaster, United Kingdom}           
\centerline{$^{23}$University of Arizona, Tucson, Arizona 85721}              
\centerline{$^{24}$Lawrence Berkeley National Laboratory and University of    
                  California, Berkeley, California 94720}                     
\centerline{$^{25}$University of California, Davis, California 95616}         
\centerline{$^{26}$California State University, Fresno, California 93740}     
\centerline{$^{27}$University of California, Irvine, California 92697}        
\centerline{$^{28}$University of California, Riverside, California 92521}     
\centerline{$^{29}$Florida State University, Tallahassee, Florida 32306}      
\centerline{$^{30}$University of Hawaii, Honolulu, Hawaii 96822}              
\centerline{$^{31}$Fermi National Accelerator Laboratory, Batavia,            
                   Illinois 60510}                                            
\centerline{$^{32}$University of Illinois at Chicago, Chicago,                
                   Illinois 60607}                                            
\centerline{$^{33}$Northern Illinois University, DeKalb, Illinois 60115}      
\centerline{$^{34}$Northwestern University, Evanston, Illinois 60208}         
\centerline{$^{35}$Indiana University, Bloomington, Indiana 47405}            
\centerline{$^{36}$University of Notre Dame, Notre Dame, Indiana 46556}       
\centerline{$^{37}$Iowa State University, Ames, Iowa 50011}                   
\centerline{$^{38}$University of Kansas, Lawrence, Kansas 66045}              
\centerline{$^{39}$Kansas State University, Manhattan, Kansas 66506}          
\centerline{$^{40}$Louisiana Tech University, Ruston, Louisiana 71272}        
\centerline{$^{41}$University of Maryland, College Park, Maryland 20742}      
\centerline{$^{42}$Boston University, Boston, Massachusetts 02215}            
\centerline{$^{43}$Northeastern University, Boston, Massachusetts 02115}      
\centerline{$^{44}$University of Michigan, Ann Arbor, Michigan 48109}         
\centerline{$^{45}$Michigan State University, East Lansing, Michigan 48824}   
\centerline{$^{46}$University of Nebraska, Lincoln, Nebraska 68588}           
\centerline{$^{47}$Columbia University, New York, New York 10027}             
\centerline{$^{48}$New York University, New York, New York 10003}             
\centerline{$^{49}$University of Rochester, Rochester, New York 14627}        
\centerline{$^{50}$State University of New York, Stony Brook,                 
                   New York 11794}                                            
\centerline{$^{51}$Brookhaven National Laboratory, Upton, New York 11973}     
\centerline{$^{52}$Langston University, Langston, Oklahoma 73050}             
\centerline{$^{53}$University of Oklahoma, Norman, Oklahoma 73019}            
\centerline{$^{54}$Brown University, Providence, Rhode Island 02912}          
\centerline{$^{55}$University of Texas, Arlington, Texas 76019}               
\centerline{$^{56}$Texas A\&M University, College Station, Texas 77843}       
\centerline{$^{57}$Rice University, Houston, Texas 77005}                     
\centerline{$^{58}$University of Washington, Seattle, Washington 98195}       
}                                                                             

\title{Search for $R$-parity Violation in Multilepton Final States 
      in $p\bar{p}$ Collisions at 
       $\sqrt{s}=1.8$~TeV}

\maketitle

\begin{abstract}
 The result of a search for gaugino pair production
  with a trilepton signature is reinterpreted 
 in the framework of
 minimal supergravity ($m$SUGRA) with $R$-parity violation via leptonic
  $\lambda$ Yukawa couplings. The search used 95 pb$^{-1}$ of $p\bar{p}$ collisions
 at $\sqrt{s}=1.8$ TeV recorded by the D\O \ detector at the Fermilab
Tevatron.
A large domain of the $m$SUGRA parameter space is excluded
for $\lambda_{121}, \ \lambda_{122} \ge 10^{-4}$.
\end{abstract}


 Supersymmetry (SUSY) is one of the possible extensions of the standard model (SM).
 For each SM particle there is a hypothesized
 supersymmetric partner with spin differing by 1/2-integer. 
 Most searches for supersymmetric particles assume 
 conservation of $R$-parity, $R_p$, a multiplicative quantum number defined as 
 $(-1)^{3N_B+N_L+2S}$, where $N_B$ is the baryon number, $N_L$ is the lepton 
 number, and $S$ is the spin quantum number~\cite{Fayet}. However, SUSY does not require 
 $R$-parity conservation. In particular, the lightest supersymmetric
 particle (LSP) can decay into a purely leptonic state
 due to the presence of an $R_p$- and $N_L$-violating term in the
 supersymmetric potential,  $\lambda_{ijk}L_iL_jE_k^C$,
 where $L_i$ and $E_k$ are isodoublet and isosinglet supersymmetric lepton fields,
 respectively (the superscript $C$ indicates charge conjugation). 
 The indices $i,j,k$ run over the three lepton generations
 and the potential is antisymmetric for the indices $i$ and $j$.
 Current upper limits on $R$-parity violating SUSY Yukawa couplings, $\lambda_{ijk}$,
 are of the order of 
 $\approx 10^{-2}$~\cite{Rlimits}.  If these couplings are not
 vanishingly small, an 
 enhancement is expected in the number of produced
 multilepton events.
 
In this paper, we reinterpret the result of a previous search by the D\O\ 
collaboration for gaugino
pair production in multilepton channels~\cite{gaugino-pair}.  
We use the
minimal low-energy supergravity model~\cite{SUGRA_rev,SUGRA_pap} ($m$SUGRA) 
as a starting point, and add non-vanishing $\lambda_{ijk}$ couplings.
 The $m$SUGRA model has four continuous parameters and one discrete parameter:  
 $m_0$ --- the universal scalar mass, $m_{1/2}$ --- the universal gaugino mass,
 $A_{0}$  --- the common trilinear interaction term, tan$\beta$ --- the ratio 
 of the vacuum expectation values of the two Higgs fields, and the sign 
 of $\mu$ --- the Higgsino mass parameter.
 The mass spectrum of the SUSY partners at the electroweak scale and their
 decay branching ratios are obtained from the above parameters by solving 
 a set of renormalization group equations using the 
 program {\sc isajet}~\cite{isajet}. 
 Present limits on the $ \lambda_{ijk}$ Yukawa couplings~\cite{Rlimits} 
 imply that this mass spectrum is the
 same as for the case of conserved $R$-parity. 
 In this analysis we consider only parameter regions with a
 neutralino ($\tilde\chi^0_1$) as LSP.

 The CDF and D\O\  collaborations have previously reported on searches for
$R$-parity violation in the di-electron+jets channels~\cite{CDF-diel,D0-diel}.
They assumed an $R_p$- and $N_L$-violating supersymmetric potential 
 term $\lambda'_{ijk}Q_iL_jD_k^C$,
 where $Q_i$ and $D_k$ are isodoublet and isosinglet supersymmetric quark fields,
 respectively.  
Some regions of the $m$SUGRA parameter space
 are excluded by non-observation of SUSY or Higgs particles at 
the CERN $e^+e^-$ collider (LEP2):
 the present limit on the mass of the lightest
 neutral SUSY Higgs boson (88.3 GeV~\cite{m-H_lim}) implies that 
 tan$\beta \le 2$ is  excluded, independent of the other parameters.
 At higher tan$\beta$,  part of the parameter
 space is excluded by the lower limit on the 
 $\tilde\chi^0_1$ mass~\cite{m-chi_lim} obtained assuming $R$-parity 
 violation through $\lambda$ couplings. 

 The event selection and background estimations used in this work are discussed 
 in the above-mentioned D\O \ search~\cite{gaugino-pair}. 
 Four different final states were considered: 
 $eee$, $ee\mu$, $e\mu\mu$, and $\mu\mu\mu$, requiring at least three electrons, 
 two electrons and a muon, two muons and an electron, or three muons, 
 in the respective channels.  No acceptable events were found. 
 The result is summarized in Table~\ref{tb:3l}. 
 The corresponding selection criteria (including the triggers) are detailed 
 in Ref.~\cite{gaugino-pair}. We consider these selection criteria adequate for
 the present analysis. 

 Our search is most sensitive
 to decays with highest electron and muon 
 multiplicity, i.e., those with no $\tau$ lepton among
 the decay products of the LSP. The detection efficiency
 is highest, especially for the case of $\lambda_{121}$, when electrons 
 dominate. On the other hand, couplings $\lambda_{133}$ and $\lambda_{233}$
 correspond to decays with least sensitivity,
 because the number of $\tau$ leptons is highest.
 We limit ourselves to the three extreme cases: 
 $\lambda_{121}$, $\lambda_{122}$ and  $\lambda_{233}$.

We generate Monte Carlo (MC) events with
all possible production and decay modes of SUSY particles 
assuming the $m$SUGRA model using {\sc isajet}~\cite{isajet} 
with $R$-parity violation~\cite{isajetR,susygen}.   
 We apply the same selection criteria 
 as used in \cite{gaugino-pair}  to these generated events, 
and calculate all signal efficiencies.


 Detector response is modeled using a parameterized, fast,
 particle-level simulation of isolated electrons, photons, and both 
isolated and non-isolated
muons. The model contains jet reconstruction 
and a simulation of the missing transverse
energy in an event. 
 Lepton acceptance criteria include the loss of electrons in
 the region between the central and end  cryostats of the calorimeter
 $(1.2 \le |\eta| \le  1.4)$, and a lookup table of the muon efficiency 
 as a function of $\eta$ and $\phi$~\cite{SGlenn_thesis,muon_acc_table},
where $\eta$ and $\phi$ are the pseudorapidity and the azimuthal angle
of the lepton, respectively.
The parameters of the program are tuned
so that the total acceptance, $\epsilon^{\rm total}$,
and the shapes
of the missing transverse energy distributions and
charged lepton $\eta$, $\phi$ and transverse
energy distributions agree with detailed simulation
based on {\sc geant}~\cite{GEANT,d0geant}. The total acceptance
includes the geometrical acceptance, 
efficiency factors for the trigger, track reconstruction, 
and lepton identification. It depends mainly
on the type of coupling and on the value of $m_{1/2}$.
In the vicinity of the exclusion contour,
the typical values are 
20\%, 10\%,  and 0.3\% for $\lambda_{121}$, 
$\lambda_{122}$, and $\lambda_{233}$, respectively.
$\epsilon^{\rm total}$ 
decreases with decreasing $m_{1/2}$, mainly because
the masses of the gauginos decrease and the energies
of their decay products fall below the detection threshold.

Our 95\% C.L. exclusion contours are based on a
Bayesian approach~\cite{Bayesian,D0Limit}.
For each point in the $(m_0,m_{1/2})$ plane, we calculate
a 95\% C.L. upper limit on the cross section. 
The excluded region is determined from the intersection of this 
surface with the corresponding cross section predicted  by 
{\sc isajet}.
In this calculation, we  use as input
the total integrated luminosities, 
and the uncertainties in the
numbers of background events 
(cf. Table~\ref{tb:3l}) and 
in $\epsilon^{\rm total}$.
The latter includes the sta\-tistical error,  
an overall 10\% systematic error
in the MC simulation, and
the error on efficiency factors for the trigger, track reconstruction, 
and lepton identification, 
determined through independent measurements described in
Ref.~\cite{gaugino-pair}.
Their values are between 10\% and 20\%, and depend on the event category
(and therefore on the $\lambda_{ijk}$ coupling)
and to a lesser extent on event kinematics 
(e.g., on supersymmetric particle masses).
Finally, we include a 10\% uncertainty on the theoretical
cross section, due to e.g., 
the choice of parton distribution function.
 
\begin{figure}[htbp]
  \epsfysize=7in\epsfbox{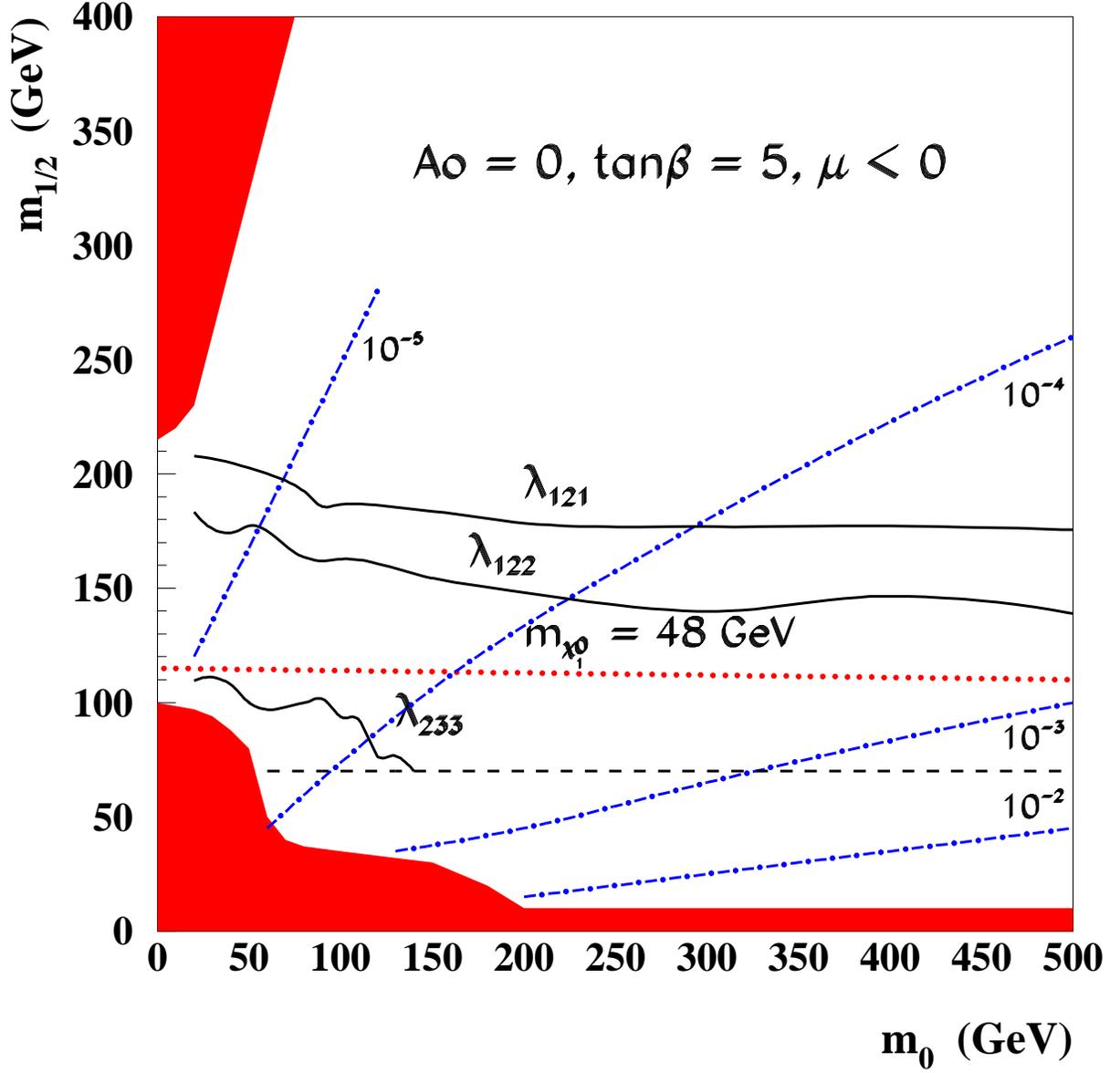}
\caption{\small Exclusion contours at 95\% C.L. limits for tan$\beta$ = 5, $\mu<$0,
for the case of finite $\lambda_{121}$, $\lambda_{122}$ and $\lambda_{233}$ 
couplings. For the explanation of the different
curves, see the text.}
\label{fg:m12_vs_m0_-5}
\end{figure}

\begin{figure}[htbp]
  \epsfysize=7in\epsfbox{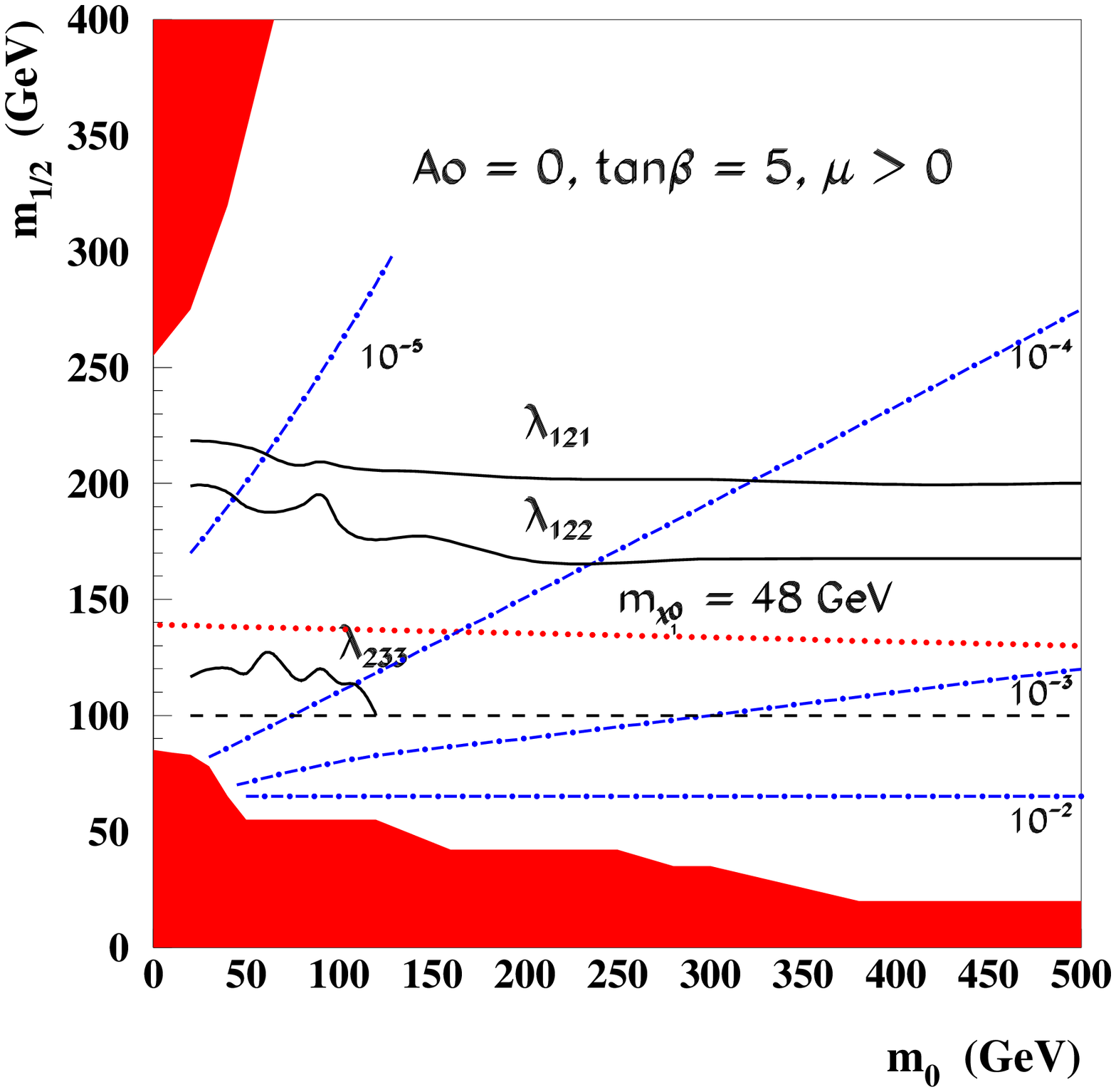}
\caption{\small Exclusion contours at 95\% C.L. limits for tan$\beta$ = 5, $\mu>$0,
for the case of finite $\lambda_{121}$, $\lambda_{122}$ and $\lambda_{233}$ 
couplings.}
\label{fg:m12_vs_m0_5}
\end{figure}

\begin{figure}[htbp]
  \epsfysize=7in\epsfbox{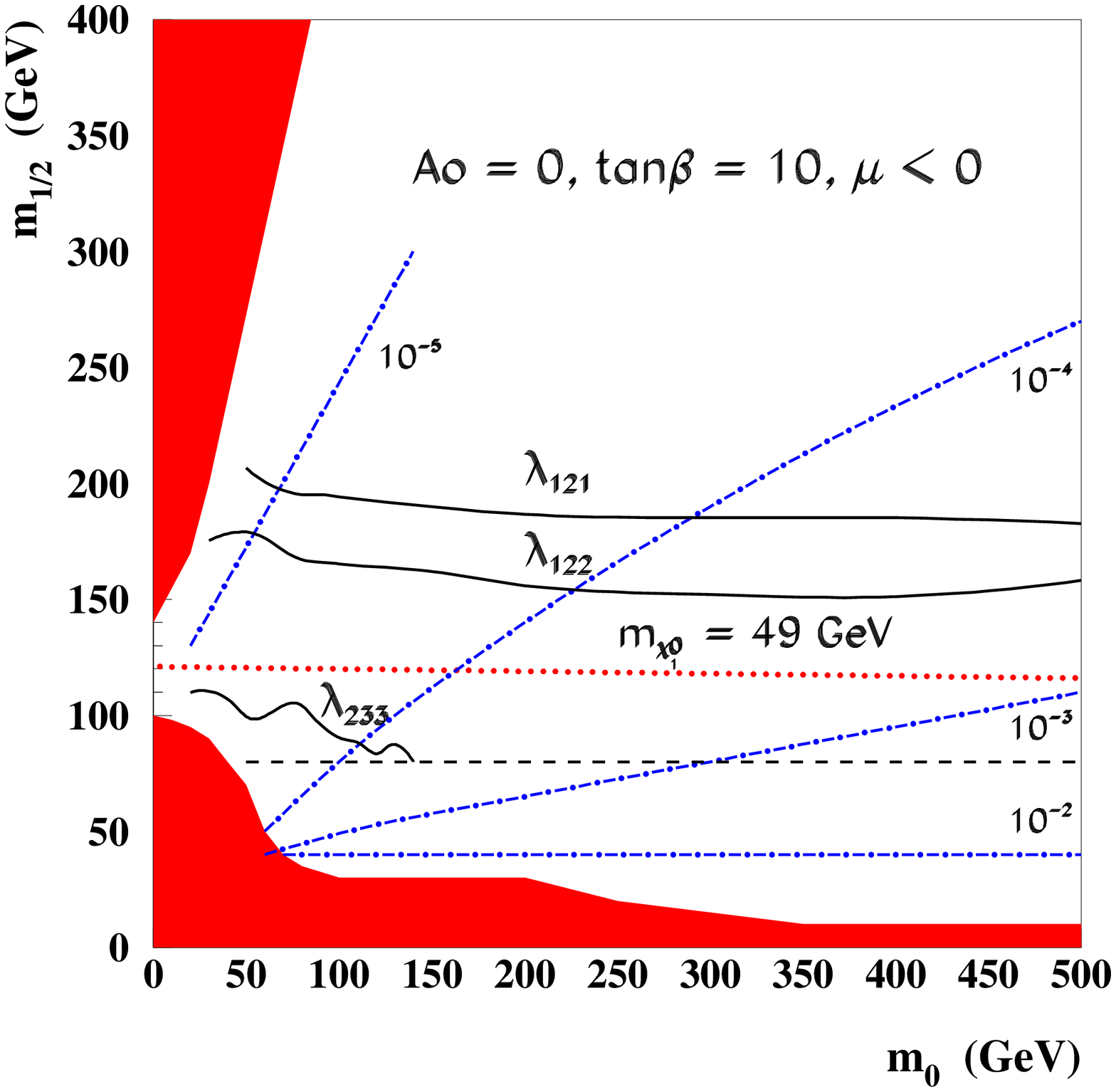}
\caption{\small Exclusion contours at 95\% C.L. limits for tan$\beta$ = 10, $\mu<$0,
for the case of finite $\lambda_{121}$, $\lambda_{122}$ and $\lambda_{233}$ 
couplings.}
\label{fg:m12_vs_m0_-10}
\end{figure}

\begin{figure}[htbp]
  \epsfysize=7in\epsfbox{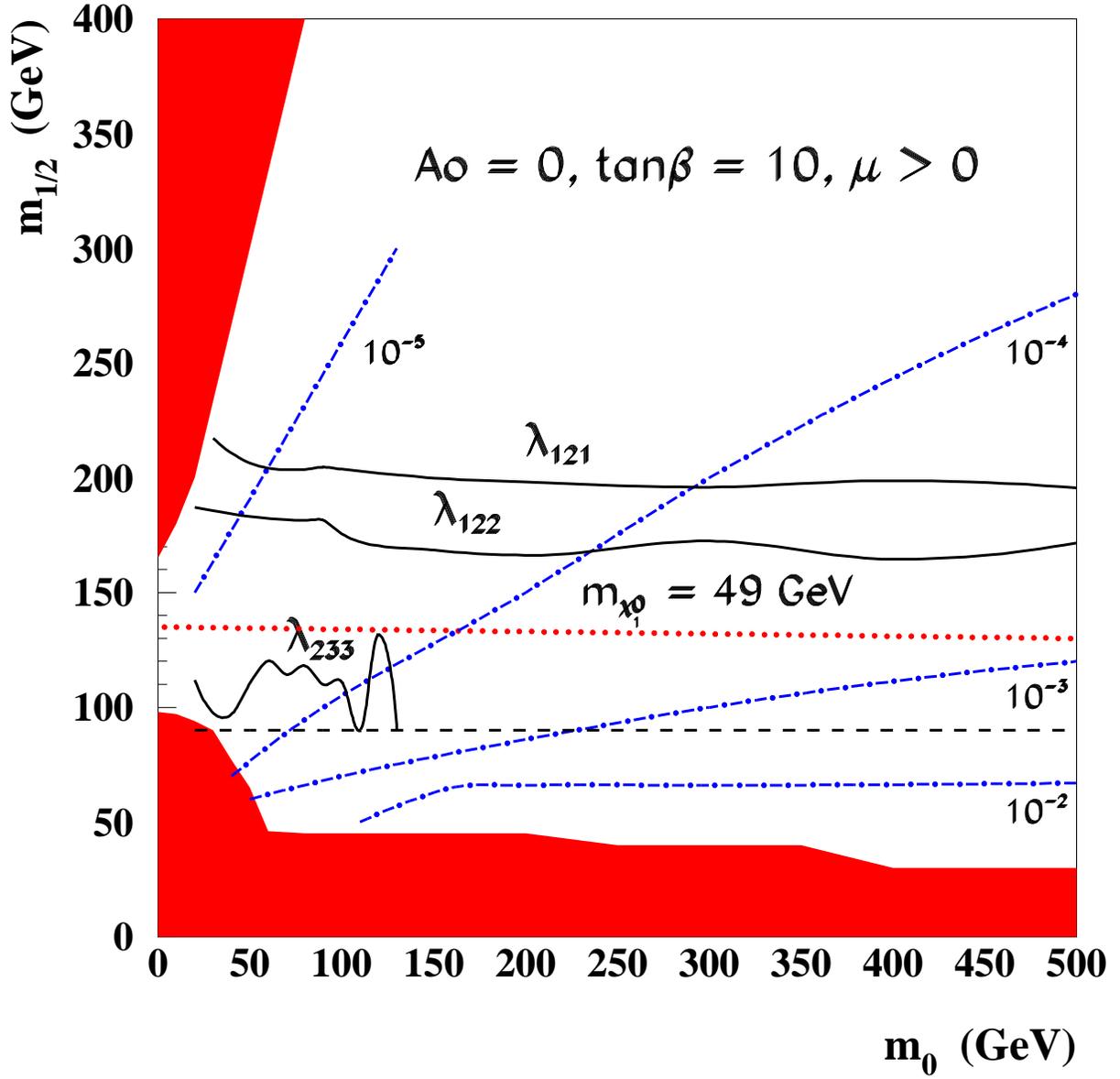}
\caption{\small Exclusion contours at 95\% C.L. limits for tan$\beta$ = 10, $\mu>$0,
for the case of finite $\lambda_{121}$, $\lambda_{122}$ and $\lambda_{233}$ 
couplings.}
\label{fg:m12_vs_m0_10}
\end{figure}

Figures \ref{fg:m12_vs_m0_-5} through \ref{fg:m12_vs_m0_10}
show, respectively, the exclusion regions in the ($m_0,m_{1/2}$) plane
for the three chosen coupl\-ings, for 
tan$\beta$ = 5 and 10, and for both signs of $\mu$.
 \ Since the characteristics of SUSY signatures at hadron colliders 
 are rather insensitive to values of $A_{0}$~\cite{A0},
we have fixed the value of $A_0$ to zero.
The dashed line indicates the limit of our sensitivity
in $m_{1/2}$ for the least favorable case,
i.e., for the coupling of $\lambda_{233}$,
where $\epsilon^{\rm total} < 10^{-4}$.
The exclusion regions correspond to the spaces
below the solid lines labelled with the coupling types, and above
the higher of the dashed line and the dash-dotted curves 
specifying the numerical values of $\lambda$.
In the regions beyond the dash-dotted curves, the average
decay length of the LSP calculated for the
value of the coupling indicated on the curve,
is less than 1 cm. 
Since efficiency studies for high impact parameter tracks
have not been done, we conservatively restrict the
present stu\-dy to decay lengths less than 1 cm.
Thus, for example, the region between curves labelled
with $\lambda_{121}$ and $10^{-3}$ is excluded if $\lambda_{121}>10^{-3}$.
The shaded areas indicate the regions
where there is no electroweak symmetry breaking
or where the LSP is not the lightest neutralino.
Finally, we also show limits
corresponding to the present lower limit on
the $\tilde\chi^0_1$ mass (dotted line),
which exclude the regions below.
The wiggles on the $\lambda_{233}$ curves are due to
statistical fluctuations and to the 
10 GeV spacing between neighboring $m_0$ points used to
calculate the curves.


In conclusion, we have reinterpreted the result of a
search for trilepton events in terms of possible $R$-parity
violation in decays of the LSP. 
We have found that a large domain of $m$SUGRA parameter
space can be excluded,
provided that $R$-parity breaking is achieved by 
lepton-number non-conservation
with $\lambda_{121}$ or $\lambda_{122}$ couplings greater than $\approx 10^{-4}$.
The region of sensitivity extends beyond that presently excluded by 
LEP experiments~\cite{m-H_lim,m-chi_lim}. For $\lambda_{233}$, where our experiment
is least sensitive, only a very limited
domain of parameter space can be excluded, and this region is already
excluded by LEP.
The excluded values of $m_{1/2}$
depend mainly on the type of coupling, and much less
on the values of other parameters. 
In particular, the excluded region is slightly larger
for $\mu>$0 than for $\mu<$0, and is almost
independent of tan$\beta$.

%
We thank the staffs at Fermilab and at collaborating institutions 
for contributions to this work, and acknowledge support from the 
Department of Energy and National Science Foundation (USA),  
Commissariat  \` a L'Energie Atomique and
CNRS/Institut National de Physique Nucl\'eaire et 
de Physique des Particules (France), 
Ministry for Science and Technology and Ministry for Atomic 
   Energy (Russia),
CAPES and CNPq (Brazil),
Departments of Atomic Energy and Science and Education (India),
Colciencias (Colombia),
CONACyT (Mexico),
Ministry of Education and KOSEF (Korea),
CONICET and UBACyT (Argentina),
A.P. Sloan Foundation,
and the Humboldt Foundation.

\begin{table}[htbp]
\begin{tabular}{l|c|c|c|c} 
Event categories &  $eee$  &  $ee\mu$  &  $e\mu\mu$  &  $\mu\mu\mu$ \\ \hline 
$\cal L_{\mbox{int}}$  ($\mbox{pb}^{-1}$) &  $98.7\pm 5.2$  &  $98.7\pm 5.2$  &  $93.1\pm 4.9$  &  $78.3\pm 4.1$ \\ \hline 
Observed events  &  0  &  0 &  0  &  0 \\ \hline 
Background events  & $0.34\pm 0.07$  &  $0.61\pm 0.36$  &  $0.11\pm 0.04$  &  $0.20\pm 0.04$ \\ 
\end{tabular}
\caption{\small The result of the search for a trilepton signature at D\O \ [3]. }
\label{tb:3l}
\end{table}

\begin{thebibliography}{99}
%
%
 
  \vskip 0.25cm


\bibitem{Fayet} G.R.~Farrar and P.~Fayet, Physics Lett. B {\bf 76}, 575 (1978).

\bibitem{Rlimits} H.~Dreiner, {\sl An Introduction to Explicit $R$ Parity Violation},
published in {\sl Perspectives in Supersymmetry},
edited by G.L.~Kane, (World Scientific, 1998);
R.~Barbier {\sl et al.}, {\sl Report of the group on the 
$R$-parity violation}, hep-ph/9810232.

\bibitem{gaugino-pair} D\O \ Collaboration, B.~Abbott {\sl et al.},  
Phys. Rev. Lett. {\bf 80}, 1591 (1998).

\bibitem{SUGRA_rev} For reviews see: H.P.~Nilles, Phys. Rep. {\bf 111}, 1 (1984); 
    H.E.~Haber and G.L.~Kane, {\sl ibid.} {\bf 117}, 75 (1985). 

\bibitem{SUGRA_pap} L.~Alvarez-Gaume, J.~Polchinski and M.B.~Wise, 
                     Nucl. Phys. {\bf B221}, 495 (1983); 
L.~Iba$\tilde{\mbox{n}}$ez, Physics Lett. B {\bf 118}, 73 (1982);  
J.~Ellis, D.V.~Nanopoulos and K.~Tamvakis, Physics Lett. B {\bf 121}, 123 (1983);  
K.~Inoue {\sl et al.}, Prog. Theor. Phys. {\bf 68}, 927 (1982); 
A.H.~Chamseddine, R.~Arnowitt and P.~Nath, Phys. Rev. Lett. {\bf 49},  970 (1982).


\bibitem{isajet} F.~Paige and S.~Protopopescu, in {\sl  Supercollider Physics},
p. 41, edited by D.~Soper, (World Scientific, 1986); 
H.~Baer, F.~Paige, S.~Protopopescu and X.~Tata, in {\sl  Proceedings of the
Workshop of Physics at Current Accelerators and Supercolliders}, edited by J.~Hewett,
A.~White and D.~Zeppenfeld, (Argonne National Laboratory, 1993). 
We used version V7.29.


\bibitem{CDF-diel}  CDF Collaboration, F.~Abe {\sl et al.}, Phys. Rev. Lett {\bf 83}, 2133 (1999).


\bibitem{D0-diel} D\O \ Collaboration, 
   B.~Abbott {\sl et al.}, Phys. Rev. Lett. {\bf 83}, 4476 (1999).

\bibitem{m-H_lim} ALEPH, Delphi, L3 and OPAL Collaborations, CERN-EP-2000-055  (2000),
submitted to ``Rencontres de Moriond'', Les Arcs, France, March 11-25, 2000.


\bibitem{m-chi_lim} ALEPH Collaboration, R.~Barate {\sl et al.}, Eur. Phys. J. C {\bf4}, 433 (1998); 
Delphi Collaboration, P.~Abreu {\sl et al.}, CERN-EP/99-49 (1999),
submitted to Eur. Phys. J. C.;
L3 Collaboration, M.~Acciari {\sl et al.}, Physics Lett. B {\bf 459}, 283 (1999); 
OPAL Collaboration, G.~Abbiendi {\sl et al.}, CERN-EP/99-123 (1999), submitted to Eur. Phys. J. C.


\bibitem{isajetR} A.~Mirea, Ph.D. thesis, Universit\'e de la M\'editerran\'ee,
  Marseille, France (1999) (unpublished).


\bibitem{susygen} S.~Katsanevas and P.~Morawitz,  Comput. Phys. Commun., {\bf 112},
227 (1998).


\bibitem{SGlenn_thesis}S.~Glenn, Ph.D thesis, University of California at Davis, (1996)
(unpublished).

\bibitem{muon_acc_table} D\O \ Collaboration, B.~Abbott {\sl et al.}, 
Phys. Rev. D {\bf 61}, 032004 (2000).

\bibitem{GEANT}R.~Brun and F.~Carminati, CERN Program Library Long Writeup W5013, 1993
(unpublished).


\bibitem{d0geant} D\O \ Collaboration, S.~Abachi {\sl et al.}, Phys. Rev. Lett. {\bf 74},
2632 (1995);  D\O \ Collaboration, B.~Abbott {\sl et al.}, Phys. Rev. D {\bf 58}, 052001 (1998).


\bibitem{Bayesian} 
H.~Jeffreys, {\sl Theory of Probability} (Clarendon Press, 
Oxford, 1961), p. 115;
G.~D'Agostini,  {\sl Bayesian reasoning in high energy physics: Principles
and applications}, CERN 99-03.

\bibitem{D0Limit} I.~Bertram  {\sl et al.}, FERMILAB-TM-2104 (unpublished).

\bibitem{A0} I.~Hinchliffe {\sl et al.},
Phys. Rev. D {\bf 55}, 5520 (1997).






\end{thebibliography}
\end{document}